\documentclass[
 aps,
 prl,
 twocolumn,
 showpacs,
 showkeys,
 amsfonts,
 amsmath,
 amssymb,
 superscriptaddress
 ]{revtex4-1}

 \usepackage{amsmath}    
 \usepackage{url}
 \usepackage{graphicx}   

\begin{document}


\title{Neutrino Velocity Anomalies: A Resolution without a Revolution}
\author{Dmitry V. Naumov}
\email{dnaumov@jinr.ru}
\affiliation{Dzhelepov Laboratory of Nuclear Problems, Joint Institute for Nuclear Research, RU-141980 Dubna, Russia}
\author{Vadim A. Naumov}
\email{vnaumov@theor.jinr.ru}
\affiliation{Bogoliubov Laboratory of Theoretical Physics, Joint Institute for Nuclear Research, RU-141980 Dubna, Russia}

\date{\today}

\begin{abstract}
We argue that the neutrino advance of time observed in MINOS and OPERA experiments can be explained
in the framework of the standard relativistic quantum theory as a manifestation of the large effective
transverse size of the eigenmass neutrino wavepackets. 
\end{abstract}

  \pacs{14.60.Lm, 14.60.Pq, 13.15.+g}

  \keywords{Neutrino;
            Charged currents;
            relativity;
            wavepacket;
           }

\maketitle

\section{Introduction}

Recently, OPERA Collaboration reported that neutrinos from CERN arrive to the 
Gran Sasso Underground Laboratory by $(60.7 \pm 6.9_{\text{stat.}} \pm 7.4_{\text{sys.}})$~ns
earlier than expected for almost massless particles~\cite{Autiero:2011hh}.
MINOS Collaboration also observed~\cite{Adamson:2007zzb} an earlier arrival
of neutrinos from FNAL to the Soudan Underground Laboratory by
$(126 \pm 32_{\text{stat.}} \pm 64_{\text{sys.}})$~ns (68\% C.L.).
The experiments have similar distances of about $730$ km between the neutrino
production and detection regions but different mean neutrino energies
(17 and 3 GeV for, respectively, CNGS and NuMI beams) and different neutrino
flavor compositions of the beams.
These remarkable results, being interpreted in terms of neutrino velocity $v_\nu$,
suggest a superluminal motion of neutrinos with~\cite{Footnote1} 
\[
v_\nu = 1+\left\{
\begin{aligned}
(5.1  \pm 2.9)  \times 10^{-5} & \enskip \text{(MINOS)}, \\
(2.48 \pm 0.41) \times 10^{-5} & \enskip \text{(OPERA)}.
\quad
\end{aligned}
\right.
\]
At the first blush this interpretation breaks the relativity principle -- one of the basis
of modern physics. We will however argue that an earlier arrival of neutrinos could be
understood without any violation of relativity, causality and other fundamental
physical concepts and is just a manifestation of quantum nature of neutrino.
Namely, we will try to demonstrate that the observed effect can be at least qualitatively
explained by taking into account that the quantum states of neutrinos with definite (small)
masses are described by the relativistic wavepackets having a finite and in fact very large
effective transversal size.
Necessity of a wavepacket description of neutrino propagation in vacuum and matter
is now well understood (while not yet commonly accepted) in the theory of neutrino flavor
transitions (``oscillations'') based on quantum mechanics or quantum field theory.


\section{Neutrino wavepacket}

Any wavepacket can be conventionally characterized by a most probable 4-momentum,
4-coordinate, and a set of parameters governing the shape of the packet in the
phase space. 
Apparently, a spherically symmetric wavepacket with an effective spatial ``size'' 
$\sigma_x$ and momentum ``width'' $\sigma_p\sim1/\sigma_x$ in its rest-frame 
becomes asymmetrical if it is boosted with a Lorentz factor $\varGamma\gg1$.
The wavepacket spatial size in the boost direction shrinks as $\sigma_x/\varGamma$
remaining unchanged in the transverse plane.
The momentum width increases in the boost direction as 
$\sigma_p\varGamma$ remaining the same in the transverse plane. In our previous 
paper~\cite{Naumov:2010um} we developed a covariant field-theoretical approach
to neutrino oscillations which operates with the relativistic wavepackets describing
initial and final states of particles involved into the production and detection
of neutrino. The neutrino in this approach is described as a virtual mass eigenfield
travelling between the macroscopically separated production and detection
vertices of Feynman graphs. Thus we make no any assumption about its wavefunction.
Instead, within our formalism we compute the neutrino wavefunction which turns out 
to be a wavepacket with spatial and momentum widths defined and functionally dependent
on those of the particles involved into the neutrino production and detection subprocesses. 
Explicitly, up to a coordinate independent spinor factor, the effective (outgoing)
neutrino wavefunction reads~\cite{Footnote2}
\begin{equation}
\label{psi_AsymptoticExpansion_lowest_LF_b}
\psi_{\nu}^* 
= e^{iE_{\nu}\left(x_0-\mathbf{v}_{\nu}\mathbf{x}\right)-\sigma_{\nu}^2\varGamma_\nu^2\left(\mathbf{x}_{\parallel}
  -\mathbf{v}_{\nu}x_0\right)^2-\sigma_{\nu}^2\mathbf{x}_{\perp}^2}.
\end{equation}
Here $x_0$ is the time, $\mathbf{x}_{\parallel}$ and $\mathbf{x}_{\perp}$ are, respectively, the longitudinal
and transverse (relative to the mean velocity vector $\mathbf{v}_{\nu}=\mathbf{p}_{\nu}/E_\nu$) spatial coordinates
of the geometric center of the neutrino packet ($x_0$ and $\mathbf{x}_{\parallel}+\mathbf{x}_{\perp}$ form a 4-vector);
$E_\nu=\sqrt{\mathbf{p}_{\nu}^2+m_{\nu}^2}$, and $m_\nu$ is the mass of the neutrino mass eigenfield.
In the most general case, the ``spread'' $\sigma_{\nu}$ is a Lorentz invariant function of the most probable 3-momenta
$\mathbf{p}_\varkappa$, masses $m_\varkappa$, and momentum spreads $\sigma_\varkappa=\text{const}$ of \emph{all}
external in and out particles $\varkappa$ involved into the neutrino production-detection process which are
described as asymptotically free relativistic wavepackets.
It is shown in Ref.~\cite{Naumov:2010um} that the center of any external wavepacket moves \emph{in the mean} along
the classical trajectory $\langle\mathbf{x}_{\varkappa}\rangle = \widetilde{\mathbf{x}}_{\varkappa}+\mathbf{v}_{\varkappa}x_{\varkappa}^0$
conserving energy, momentum and effective volume ($\propto1/\sigma_\varkappa^3$);
under certain conditions the packets remain stable (nondiffluent) during the times much longer than their mean lifetimes 
(in case of unstable particles) or the mean time between the two successive collisions in the relevant ensemble
(in case of stable particles). 

Considering that the two-body decays of pions and kaons 
are the main processes of neutrino  production in the MINOS and OPERA experiments, we can neglect
the contributions into $\sigma_{\nu}$ coming from the particles, interacting with neutrinos in the detector
(reasonably assuming that their 4-momentum spreads are much larger than $\sigma_\pi$, $\sigma_K$ and $\sigma_\mu$).
With this simplification, it can be proved that
\[
\sigma_{\nu}^2 \approx m_\nu^2\left(\frac{m_a^2}{\sigma_a^2}+\frac{m_\mu^2}{\sigma_\mu^2}\right)^{-1},
\quad
a=\pi~\text{or}~K.
\]
Then from the above-mentioned conditions of stability for the meson and muon wavepackets
it follows that $\sigma_\nu$ must satisfy the following conditions
\[
\sigma_\nu^2 \ll m^2_\nu\left(\frac{m_\mu}{\Gamma_\mu}+\frac{m_a}{\Gamma_a}\right)^{-1},
\]
where $\Gamma_a=1/\tau_a$ and $\Gamma_\mu=1/\tau_\mu$ are the full decay widths of the meson $a$ and muon.
Considering that for \emph{any} know meson $m_\mu/\Gamma_\mu \gg m_a/\Gamma_a$, we conclude that 
\begin{equation}
\label{eq:sigmap_limit}
\sigma_\nu^2/m^2_\nu\ll\Gamma_\mu/m_\mu \approx 2.8\times10^{-18}.
\end{equation}
Therefore the neutrino momentum uncertainty is fantastically small~\cite{Footnote3}.
From~\eqref{eq:sigmap_limit} one can immediately derive the lower bounds for the effective spatial
dimensions of the neutrino wavepacket: 
\begin{gather*}
d_\perp \gg \left(\frac{0.1~\text{eV}}{m_\nu}\right)~\text{km}, \\
d_\parallel = \frac{d_\perp}{\varGamma_\nu} 
        \gg 10^{-2}\left(\frac{10~\text{GeV}}{E_\nu}\right)\left(\frac{0.1~\text{eV}}{m_\nu}\right)~\mu\text{m},
\end{gather*}
So the neutrino wavepacket appears as a huge but superfine disk of microscopic (energy dependent)
thickness in longitudinal direction, comparable with the thickness of a soap-bubble skin,
and macroscopically large (energy independent) diameter in the transverse plane~\cite{Footnote4}.

This is a key point in interpretation of the experiments with earlier arrival of neutrino signal.

\section{Qualitative estimations}

\begin{figure}[htb]
 \begin{center}
 \includegraphics[width=0.90\linewidth]{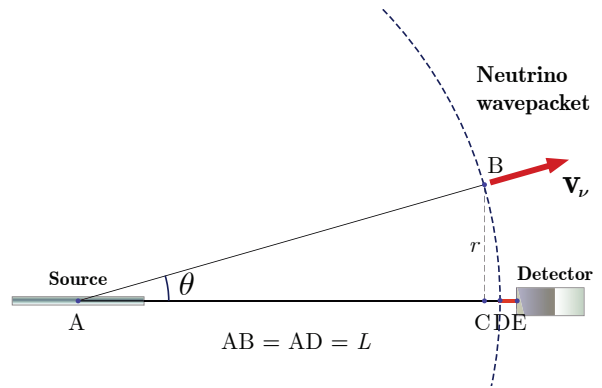}
 \caption{Neutrinos are emitted from the ``Source'' and are registered in the ``Detector''.
          The centers of the neutrino wavepackets will arrive at the points B and D
          simultaneously, while the signal from the neutrino wavepacket
          (shown as an extremely oblate spheroid) which moves under the angle $\theta=\angle\text{BAC}$
          to the beam axis will arrive earlier since $\text{DE}>0$.
          Neutrino velocity vector $\mathbf{v}_\nu$ lies in the plane of the figure.
          Proportions do not conform to reality.
          }
 \label{fig:NeutrinoWavepacket}  
 \end{center}
\end{figure}

\begin{figure}[htb]
 \begin{center}
 \includegraphics[width=0.90\linewidth]{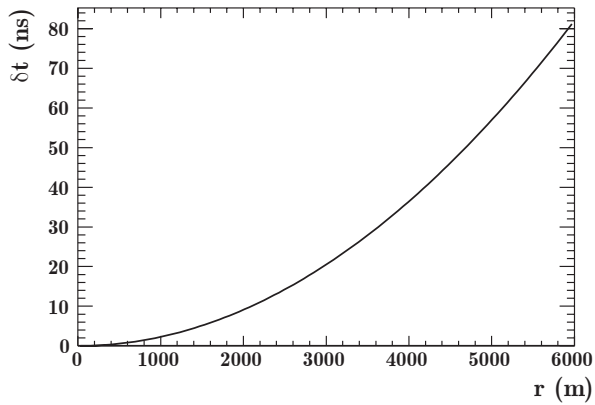}
 \caption{Advance $\delta t$ as a function of $r$.}
 \label{fig:deltaT_vs_r}  
 \end{center}
\end{figure}

In fact, neutrinos produced at accelerators arrive to the detector site widely 
distributed across the plane transverse to the beam axis. Those neutrinos which
were misaligned with the neutrino detector nevertheless do have a chance to
interact within the detector due to macroscopically large transverse size
of its wavefunction. Moreover, its interaction probability very weakly depends
on the misalignment distance $r=\text{BC}$ (see Fig.~\ref{fig:NeutrinoWavepacket})
if it is small compared to $d_\perp$.
As is seen from Fig.~\ref{fig:NeutrinoWavepacket}, the misaligned neutrinos will interact
{\em systematically earlier} than those moving along the beam axis, due the huge transverse width
of their wavefunctions. 
The school-level planimetry suggests that the advancing time is given by
\begin{equation}
\label{eq:advanceTime}
\delta t = L\left(1/\cos\theta-1\right)\approx r^2/(2L).
\end{equation}
Here we assume that (i) $1-v_\nu \lll 1$, (ii) the neutrino wavepacket effective width
is much larger than the detector dimensions, and (iii) $\theta \ll 1$.
Substituting numbers into \eqref{eq:advanceTime} one obtains Fig.~\ref{fig:deltaT_vs_r}
from which quantitative estimates for the time advance as a function of $r$ could be drawn.
For instance, a neutrino packet which moves 3~km away from the detector will come earlier
by about $20$~ns than that moving directly to the detector. \\

What is the probability to find a neutrino at a distance $r$ from the beam axis? 
This could be estimated taking into account that neutrino production is 
dominated by two-particle decays of pions and kaons. The angular distribution of
massless neutrinos from these decays is well known:
\begin{equation}
\label{eq:AngularDistribution}
\frac{dI}{d\Omega}= \frac{1-v_a^2}{4\pi(1-v_a\cos\theta)^2} \approx \frac{1}{\pi(1+\varGamma_a^2\theta^2)^2}.
\end{equation}
Here $\theta$ is the angle between the momenta of the meson $a$ and neutrino ($0\le\theta\le\pi$),
$v_a$ is the meson velocity, and $\varGamma_a=(1-v_a^2)^{-1/2}=E_a/m_a$.
The second approximate equality in Eq.~\eqref{eq:AngularDistribution} holds for small angles
and relativistic meson energies ($\theta\ll1$, $4\varGamma_a^2 \gg1$).
In the latter case, the main contribution to the neutrino event rate comes from the narrow cone
$\theta \lesssim 1/\varGamma_a$. Considering that the mean neutrino energy, $\overline{E}_\nu$,
from the muonic decay of a meson with energy $E_a$ is $\overline{E}_\nu=\varGamma_aE_\nu^{(a)}$,
where $E_\nu^{(a)}=(m_a^2-m_\mu^2)/(2m_a)$ is the neutrino energy in the rest frame
of the particle $a$, the characteristic angle can be defined as $\theta_{(a)}=E_\nu^{(a)}/\overline{E}_\nu$.

In the case of OPERA, one can (very) roughly estimate the characteristic angles for the
``low-energy'' (LE) range ($E_\nu<20$~GeV, $\overline{E}_\nu \approx 13.9$~GeV) and
``high-energy'' (HE) range ($E_\nu>20$~GeV, $\overline{E}_\nu \approx 42.9$~GeV),
assuming that the main neutrino sources in these ranges are, respectively, $\pi_{\mu2}$ and $K_{\mu2}$ decays:
\begin{equation*}
\label{theta_pi_K}
\begin{aligned}
\theta_{\text{LE}}~\gtrsim~\theta_{(\pi)}=2.1\times10^{-3}, \\
\theta_{\text{HE}}~\lesssim~\theta_{(K)}=5.5\times10^{-3}.
\end{aligned}
\end{equation*}
This provides us with an order-of-magnitude estimate of the mean values of $r$
and advancing times ${\delta}t$:
\begin{gather*}
\label{eq:r_OPERA}
r_{\text{LE}}~\gtrsim~1.7~\text{km},
\quad
r_{\text{HE}}~\lesssim~11~\text{km}; \\
 \label{eq:deltaT_OPERA}
{\delta}t_{\text{LE}}~\gtrsim~5.6~\text{ns},
\quad
{\delta}t_{\text{HE}}~\lesssim~36.7~\text{ns}.
\end{gather*}
Since the LE and HE ranges contribute almost equally to the CNGS $\nu_\mu$ beam,
there must be a definite trend towards earlier neutrino arrival to OPERA with
approximately $21$~ns mean time-shift and a ``tail'' or, better to say, ``fore''
of the same order coming from the ``edges'' of the CNGS beam.

Similar estimation for the low-energy NuMI beam at Fermilab producing neutrinos
for the MINOS experiment can be done with a better accuracy, since the $\pi_{\mu2}$ decay
is here the dominant source of neutrinos and the radial distribution of the beam is expected
to be very flat. So, by using $\overline{E}_\nu=3$~GeV we obtain
\begin{equation}
\label{eq:deltaT_MINOS}
r \approx 36.2~\text{km},
\quad
{\delta}t \approx 120.7~\text{ns}.
\end{equation}
The latter number is in surprisingly good agreement with the MINOS observation.
Obviously, MINOS should observe at the average a much earlier arrival of neutrinos,
in comparison with OPERA, because of the lower mean neutrino energy which corresponds
to a wider transverse beam distribution and hence to a larger input from the misaligned neutrinos.


\section{Numerical results}

\begin{figure}[htb]
 \begin{center}
 \includegraphics[width=0.90\linewidth]{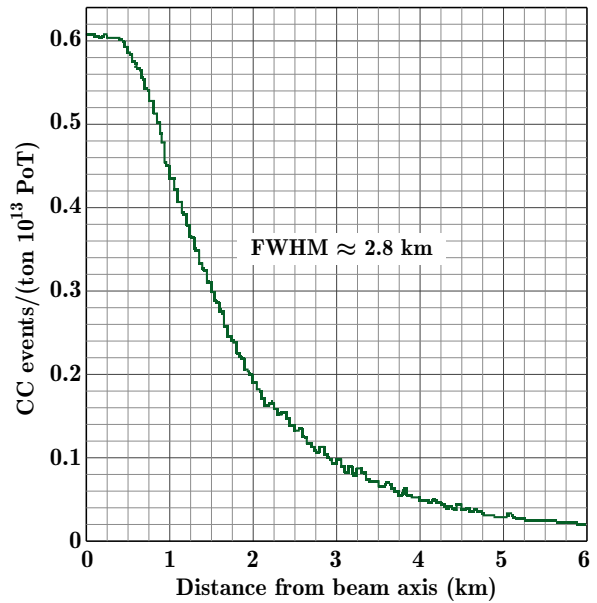}
 \caption{Probability of neutrino charged current interactions expected in OPERA as function of $r$.}
 \label{fig:flux_vs_R_OPERA}  
 \end{center}
\end{figure}

Let us reevaluate the estimates given above with a somewhat detailed but still 
simplified calculation. In particular, we could profit from the simulation of 
expected radial distribution of $\nu_\mu$ charged current (CC) events
performed by the OPERA Collaboration~\cite{NuFluxOPERA_Radial}. This distribution 
($P_{\text{\text{CC}}}(r)$) which we digitalized for our purposes is displayed in 
Fig.~\ref{fig:flux_vs_R_OPERA}. Being dominated by the $\pi_{\mu2}$ and $K_{\mu2}$
decays, the transverse beam size at Gran Sasso is of the order of kilometres
and the full width at half maximum of the distribution is about 2.8~km.

\begin{figure}[htb]
 \begin{center}
 \includegraphics[width=0.90\linewidth]{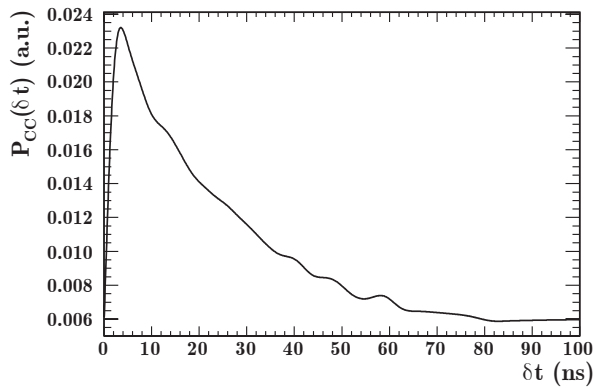}
 \caption{Advance $\delta t$ distribution expected in OPERA.}
 \label{fig:deltaT_distribution_OPERA}  
 \end{center}
\end{figure}

\begin{figure}[htb]
 \begin{center}
 \includegraphics[width=0.90\linewidth]{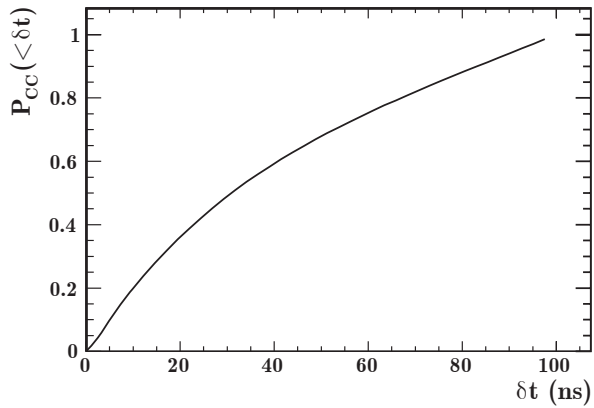}
 \caption{$P_{\text{CC}}(<\delta t)$ distribution expected in OPERA.}
 \label{fig:deltaT_int_distribution_OPERA}  
 \end{center}
\end{figure}

The distribution $P_{\text{CC}}(r)$ transformed (with help of Eq.~\eqref{eq:advanceTime})
into the $\delta t$ distribution as 
\[
P_{\text{CC}}(\delta t) = \frac{rP_{\text{CC}}\left(r\left(\delta t\right)\right)}{\displaystyle\int_0^\infty dr rP_{\text{CC}}(r)}
\] 
is shown in Fig.~\ref{fig:deltaT_distribution_OPERA}.
Its average $\langle\delta t\rangle$ is about 20~ns with similar variance and with
the tail extending up to about 100~ns.
Figure~\ref{fig:deltaT_int_distribution_OPERA} shows the integral distribution 
\[
P_{\text{CC}}\left(<\delta t\right) = \int_0^{\delta t} dt' P_{\text{CC}}(t').
\] 
Examination of Fig.~\ref{fig:deltaT_int_distribution_OPERA} suggests that all 
CC events roughly equally populate the following intervals in $\delta t$:
$(0,20)$~ns, $(20,45)$~ns, and $(45,100)$~ns.

\begin{figure}[htb]
 \begin{center}
 \includegraphics[width=\linewidth]{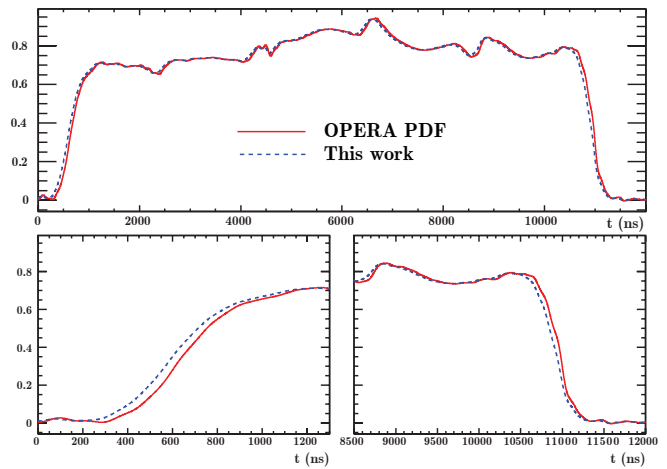}
 \caption{Top panel: time probability density function for the first beam extraction taken as an example,
          expected by the OPERA Collaboration~\cite{Autiero:2011hh} and our calculation.
          In both cases, the systematic ``instrumental'' shift is not shown since it does not change the shape of the curves.
          Bottom left and right panels: zooms of the top panel for the left and right fronts of the signal, respectively.
}
 \label{fig:shift_time_cut}
 \end{center}
\end{figure}

Finally, we compute the expected time distribution in OPERA, $g(t)$, as a convolution
of the probability density function of arrival time $f(t)$ shown in top panel of
Fig.~\ref{fig:shift_time_cut} as solid line, taking into account an earlier arrival of neutrino signal as follows:
\begin{equation*}
g(t) = \frac{\displaystyle\int_0^\infty f\left(t+\delta t(r)\right)P_{\text{CC}}(r)r dr}{\displaystyle\int_0^\infty P_{\text{CC}}(r)rdr}.
\end{equation*}
The resulting curve $g(t)$ is displayed superimposed in the top panel of Fig.~\ref{fig:shift_time_cut} by dashed line.
As is seen, on the average, it is shifted to the left by about 20~ns. However, and this is even more important,
the left front of the signal is shifted by a larger amount as it accumulates the advance effect from the total $f(t)$ distribution.
One could verify that the solid and dashed lines shown in the bottom panel of Fig.~\ref{fig:shift_time_cut} for the
left front of the OPERA time distribution is shifted by about 50~ns (up to 60~ns on the tail) to the left,
while the right front is shifted only by 20--25~ns. In other words, the impact of the misaligned neutrinos is predicted
to be asymmetric in time. 

In the likelihood fit of the time distribution performed by the OPERA Collaboration the fronts of the time distribution
statistically play the major role. Therefore, the time distortion evaluated in the present work seem to explain the OPERA
observations without any model parameter and without introducing superluminal neutrinos or other exotics.

\section{Conclusions}

Large transverse size of the neutrino wavepacket and uncollimated beam of neutrinos 
seem to explain the earlier arrival of the neutrino signal in OPERA and MINOS.
The neutrino signal is estimated to arrive in advance by about 20~ns in the mean
(with a similar variance) for OPERA and by about 120~ns for MINOS.
In the case of the OPERA experiment only this effect essentially reduces the statistical
significance of its observation. Moreover, we have evaluated the expected time distribution
of the neutrino arrival in OPERA and obtained that the left and right fronts are shifted
to the left by about 50--60~ns and 20--25~ns, respectively. This probably explains the
observed anomaly all-in-all without any exotic hypothesis, like Lorentz violation and so on.
Let us underline that in our calculations we do not use any adjustable parameter.
In the case of the MINOS experiment there is also a surprisingly good agreement between our
expectation \eqref{eq:deltaT_MINOS} and experimental result. Therefore, we argue that observations
of superluminal neutrinos by the OPERA and MINOS experiments can be treated as a manifestation
of the huge transverse size of the neutrino wavefunction.
This kind of effects could be investigated in the future experiments (in particular,
in the off-axis neutrino experiments) with more details in order to prove or disprove
our explanation.

Let us note that one should not expect an increase in the number of neutrino induced
events due to the misaligned neutrino interactions because this effect will be compensated
by the corresponding decrease of the number of aligned neutrinos.

Let us briefly discuss the situation with the observed (anti)neutrino signal from SN1987A.
A proper treatment of these neutrinos should take care about the dispersion of the neutrino
wavepackets at astronomical distances.
Deliberately neglecting the dispersion, it appears that any terrestrial detector is
sensitive only to the aligned neutrinos, since the misaligned neutrinos will have
negligible impact due to the smallness of their wavepacket transverse size relative to the
astrophysical scale of about 50~kps. Therefore, no advance signal should be expected.
However this problem is not so simple and needs in a more detailed theoretical analysis.

\begin{acknowledgments}

This work was supported by the Federal Target Program ``Scientific and scientific-pedagogical
personnel of the innovative Russia'', contract No.~02.740.11.5220.

\end{acknowledgments}


\begin{thebibliography}{5}
\bibitem{Autiero:2011hh}
  T.~Adam {\it et al.} (OPERA Collaboration),
  arXiv:1109.4897 [hep-ex].

  \bibitem{Adamson:2007zzb}
  P.~Adamson {\it et al.} (MINOS Collaboration),
  Phys.\ Rev.\ D {\bf 76}, 072005 (2007)
  [arXiv:0706.0437 [hep-ex]].

\bibitem{Footnote1}
  The MINOS Collaboration providently concludes that their measurement is consistent
  with the speed of light to less than $1.8\sigma$ and the corresponding 99\% C.L.\
  bounds on $v_\nu$ are $-2.4\times10^{-5} < v_\nu-1 < 12.6 \times 10^{-5}$.

\bibitem{Naumov:2010um}
  D.~V.~Naumov, V.~A.~Naumov,
  J.\ Phys.\ G {\bf G37}, 105014 (2010)
  [arXiv:1008.0306 [hep-ph]].

\bibitem{Footnote2}
  In fact expression~\eqref{psi_AsymptoticExpansion_lowest_LF_b} has a limited range of applicability
  which is however deliberately comprises all terrestrial neutrino experiments.

\bibitem{Footnote3}
  By the way, Eq.~\eqref{eq:sigmap_limit} explains a success of the naive quantum-mechanical approach to 
  neutrino oscillations which is explicitly based on the assumption that the massive neutrinos have
  definite 3-momenta and thus can be described by plain waves (infinitely large mathematical artifacts).

\bibitem{Footnote4}
  Note that the acceptable effective transverse dimensions of ``normal'' particle states
  (or, that is the same, the effective diameters of the packets in the intrinsic frame of reference)
  are not so huge as for neutrinos.
  For example, the limiting dimensions (minimal effective diameters of the wavepackets, necessary to keep
  them stable during the life of the particles) should be of the order of $10^{-5}$, $10^{-4}$, and $10^{-8}$~cm
  for $\pi^{\pm}$, $\mu^{\pm}$, and $\tau^{\pm}$, respectively. These dimensions are probably hard to measure
  in the current particle scattering experiments, since the wavepackets are reduced due to any measurement (interaction).
  Neutrinos provide the only (as yet) tool for such measurements, which are believed to be already realized 
  by the MINOS and OPERA experiments.

\bibitem{NuFluxOPERA_Radial} \url{http://proj-cngs.web.cern.ch/proj-cngs/Beam%20Performance/NeutrinoRadial.htm}

\end{thebibliography}
\end{document}